\documentclass[%
 reprint,
superscriptaddress,
 amsmath,amssymb,
 aps,prl,
]{revtex4-2}
\usepackage{float}
\usepackage{graphicx}% Include figure files
\usepackage{dcolumn}% Align table columns on decimal point
\usepackage{bm}% bold math
\usepackage[colorlinks=true, letterpaper=true, pdfstartview=FitV, linkcolor=red, citecolor=blue, urlcolor=blue]{hyperref}
\begin{document}

\preprint{APS/123-QED}

\title{Quantum Sensing Beyond Exceptional Points Via Hidden Symmetry-protected Vacuum-noise Fixed Point}

\author{Wencong Wang}
\affiliation{Guangdong Provincial Engineering Research Center for Optoelectronic Instrument, School of Electronic Science and Engineering (School of Microelectronics), South China Normal University, Foshan 528225, China}
\author{Yuyang Liang}
\affiliation{Key Laboratory of Atomic and Subatomic Structure and Quantum Control (Ministry of Education), Guangdong Basic Research Center of Excellence for Structure and Fundamental Interactions of Matter, School of Physics, South China Normal University, Guangzhou 510006, China}
\affiliation{Guangdong Provincial Key Laboratory of Quantum Engineering and Quantum Materials, Guangdong-Hong Kong Joint Laboratory of Quantum Matter, South China Normal University, Guangzhou 510006, China}
\author{Peng Han}
\affiliation{Guangdong Provincial Engineering Research Center for Optoelectronic Instrument, School of Electronic Science and Engineering (School of Microelectronics), South China Normal University, Foshan 528225, China}
\affiliation{Quantum Science Center of Guangdong-HongKong-Macao Greater Bay Area(Guangdong), Shenzhen, 518000, China}
\author{Xianqiu Wu}
\affiliation{Key Laboratory of Atomic and Subatomic Structure and Quantum Control (Ministry of Education), Guangdong Basic Research Center of Excellence for Structure and Fundamental Interactions of Matter, School of Physics, South China Normal University, Guangzhou 510006, China}
\affiliation{Guangdong Provincial Key Laboratory of Quantum Engineering and Quantum Materials, Guangdong-Hong Kong Joint Laboratory of Quantum Matter, South China Normal University, Guangzhou 510006, China}
\author{Dongmei Liu}
\email{dmliu@scnu.edu.cn}
\affiliation{Guangdong Provincial Engineering Research Center for Optoelectronic Instrument, School of Electronic Science and Engineering (School of Microelectronics), South China Normal University, Foshan 528225, China}
\affiliation{Quantum Science Center of Guangdong-HongKong-Macao Greater Bay Area(Guangdong), Shenzhen, 518000, China}
\author{Min Gu}
\email{mingu@m.scnu.edu.cn}
\affiliation{Key Laboratory of Atomic and Subatomic Structure and Quantum Control (Ministry of Education), Guangdong Basic Research Center of Excellence for Structure and Fundamental Interactions of Matter, School of Physics, South China Normal University, Guangzhou 510006, China}
\affiliation{Guangdong Provincial Key Laboratory of Quantum Engineering and Quantum Materials, Guangdong-Hong Kong Joint Laboratory of Quantum Matter, South China Normal University, Guangzhou 510006, China}
\affiliation{Quantum Science Center of Guangdong-HongKong-Macao Greater Bay Area(Guangdong), Shenzhen, 518000, China}

\begin{abstract} 
Exceptional-point (EP) sensing has attracted considerable interest because of its anomalous response scaling. However, recent studies have shown that the enhanced response near an EP is inevitably accompanied by amplified quantum noise, fundamentally limiting the achievable signal-to-noise ratio (SNR). Here, we propose a fundamentally different route toward non-Hermitian quantum sensing based on symmetry-protected noise suppression rather than response amplification. We develop a fully quantum continuous-variable model that unifies parity-time (PT) and anti-parity-time (APT) symmetries within a single framework. Exploiting the incompatibility between these two symmetries, we uncover a non-Hermitian Dirac eigenspectrum and reveal a hidden symmetry-protected phase transition embedded in the Hamiltonian spectrum. Remarkably, this hidden phase transition simultaneously constitutes a symmetry-protected vacuum-noise fixed point, where collective three-mode quadratures exhibit suppressed quantum fluctuations despite the absence of any anomalous spectral response. As a consequence, quantum sensing is enhanced through the suppression of excess quantum noise while maintaining a finite response sensitivity, establishing a sensing mechanism fundamentally different from conventional EP-based approaches. These results reveal an unexpected connection between hidden symmetry, quantum fluctuations, and non-Hermitian quantum metrology, and establish noise suppression as an alternative paradigm for non-Hermitian quantum sensing.
\end{abstract}

\maketitle
%%%%%%%%%%%%%%%%%%%%%%%%%%  body  %%%%%%%%%%%%%%%%%%%%%%%%%%
\section{Introduction}
%\emph{{Introduction}.---}
Non-Hermitian parity-time (PT) symmetry \cite{bender1998real} and its counterpart, anti-parity-time (APT) symmetry, have attracted extensive attention since their introduction and have been widely explored in photonics\cite{1,2,3,4}, acoustics\cite{huang2024acoustic,gu2021controlling,wu2026observation}, cold atoms\cite{jiang2019anti,peng2016anti,zhang2018non}, and many other physical platforms\cite{cao2024observation,wang2026enhancement}. Among the most intriguing phenomena in PT/APT systems are non-Hermitian exceptional points (EPs), where both eigenvalues and eigenvectors coalesce. Owing to their unique physical properties and nontrivial topological structures, EPs give rise to a variety of unconventional effects and functionalities, including asymmetric mode switching\cite{2.1,zhang2019dynamically,li2020hamiltonian}, unidirectional reflectionless transport\cite{5,6,7,li2025low}, and topological state transfer\cite{hong2025observation,zhang2025topological}. In particular, compared with conventional Hermitian sensors whose eigenvalue response scales linearly with perturbations, an $n$ th-order EP exhibits a characteristic eigenvalue splitting proportional to $\epsilon^{1/n}$, making EPs highly promising for enhanced quantum sensing applications\cite{tang2023pt,li2023stochastic}.

A long-standing objective in non-Hermitian quantum sensing is to surpass conventional precision limits under noisy environments by simultaneously improving measurement sensitivity and signal-to-noise ratio (SNR). Although EP-enhanced sensing has achieved remarkable successes in classical platforms such as electronic circuits \cite{bai2024observation,sun2024non,chen2026nonlinear,hu2025linewidth} and photonic systems\cite{ruan2025observation,luo2022quantum}, recent theoretical and experimental studies \cite{loughlin2024exceptional,almanakly2026probing,wiersig2023petermann,wang2020petermann} have revealed a fundamental limitation in passive PT-symmetric quantum sensors: the sensitivity enhancement near an EP is inevitably accompanied by amplified quantum noise, resulting in little or no net improvement in the achievable SNR. To overcome this obstacle, most existing approaches focus on modifying the scaling law of the sensing response, for example by exploiting nonlinear exceptional points\cite{bai2024observation,zheng2025noise,chen2026nonlinear} or quantum-squeezed resources\cite{wang2025quadrature,wang2026squeezing}. 
However, enhancing the response sensitivity is not the only possible route toward quantum-enhanced sensing. Within the framework of quantum metrology, the sensing performance of an observable $O$ with respect to a parameter $\theta$ can be characterized by the unitless sensitivity\cite{luo2022quantum,almanakly2026probing,xiao2024non,pezze2018quantum}
\begin{equation}
	S(O,\theta)=\frac{(\partial_{\theta}\left\langle O \right\rangle)^2}{\left\langle \Delta O^2 \right\rangle}.\label{eq:sensitivity}
\end{equation}
where $\partial_{\theta}\left\langle O \right\rangle$ denotes the response sensitivity of the expectation value of the measured observable to variations in the physical parameter $\theta$, while $\left\langle \Delta O^2 \right\rangle$ represents the noise fluctuation of the measuremet outcomes of the observable. According to the error-propagation formula, this quantity provides an experimentally accessible lower bound on the achievable sensing performance and is bounded from above by the quantum Fisher information\cite{pezze2018quantum,braunstein2005quantum}. Eq.(\ref{eq:sensitivity}) therefore indicates that the ultimate sensing performance is determined not only by the response sensitivity but also by the accompanying noise fluctuations.

In this work, we propose an alternative strategy for non-Hermitian quantum sensing by enhancing the SNR through suppressing quantum noise fluctuations, rather than through anomalous response amplification near an EP. To this end, we construct a fully quantum continuous-variable model that unifies PT and APT symmetries within a single framework. Exploiting the incompatibility between these two symmetries, we uncover a non-Hermitian Dirac eigenspectrum and demonstrate that a symmetry-protected vacuum-noise fixed point (VFP) can remain hidden within the Hamiltonian eigenspectrum under the protection of an emergent chiral symmetry. By analyzing the quantum fluctuations, we identify a set of collective three-mode quadratures governed by this hidden fixed point. Within the framework of quantum metrology, we show that these quadratures exhibit enhanced sensing performance at the symmetry-protected VFP. Unlike conventional EP-based sensing, the proposed mechanism improves the SNR through the suppression of quantum fluctuations while maintaining a finite response sensitivity, without relying on anomalous response scaling. Our work therefore establishes a fundamentally different paradigm for non-Hermitian quantum sensing and reveals a previously unexplored role of hidden symmetry-protected phase transitions in quantum-enhanced measurements. More broadly, our work reveals the previously unexplored role of VFP in quantum-enhanced measurements and provides an alternative paradigm for non-Hermitian quantum sensing.

\section{Theoretical Model}
%\emph{{Theoretical model anaysis}.---}
\begin{figure}[t]
	\centering\includegraphics[width=1\linewidth]{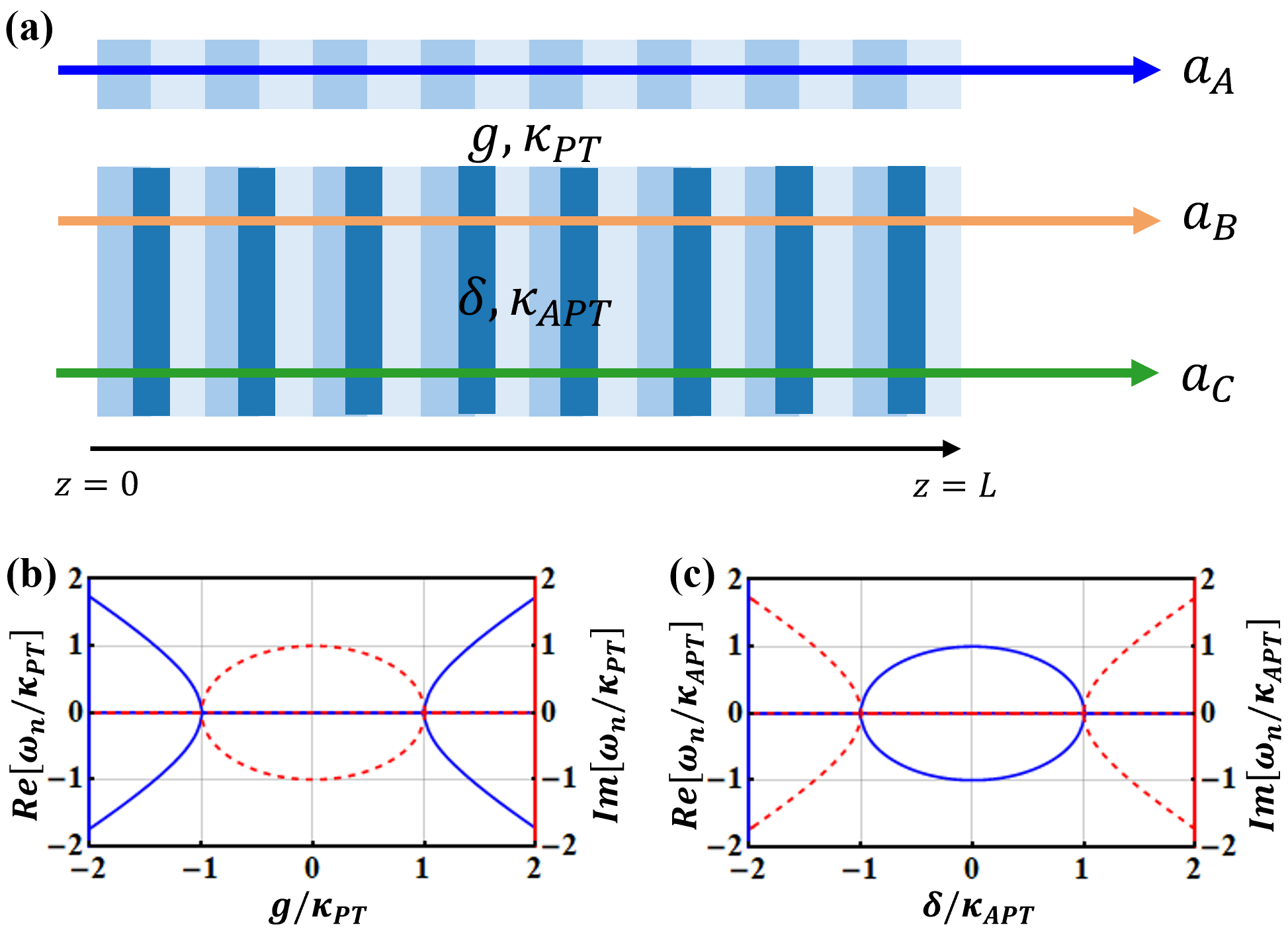}
	\caption{(a) Schematic illustration of the hybrid system formed by coupling degenerate phase-sensitive amplification and non-degenerate parametric process through linear interaction between mode $A$ of the amplification and mode $B$ of amplification and detuned non-degenerate parametric process. Real (blue solid) and imaginary (red dashed) parts of the eigenvalues $\omega_n$ for (b) the pure quadrature-PT-symmetry, ${\kappa_{APT},\delta}={0,0}$, and (c) the pure quadrature-APT-symmetry, ${\kappa_{PT},g}={0,0}$.}
	\label{fig:scheme}
\end{figure}
Before proceeding, we note that PT-symmetric systems can be realized through linear coupling between two degenerate parametric amplification processes \cite{wang2026realization,hernandez2025quantum,roy2021nondissipative}, whereas APT-symmetric systems can be implemented using detuned nondegenerate parametric amplification processes \cite{luo2022quantum,jiang2019anti,fan2020antiparity}. Inspired by this, we consider a three-mode bosonic model formed by coupling degenerate phase-sensitive amplification and detuned non-degenerate parametric process, as shown in Fig. \ref{fig:scheme}(a). We assume that modes A and B have the same wavelength. The three-mode bosonic fully quantum continuous-variable model described by the second-quantized Hermitian Hamiltonian 
\begin{equation}
	\begin{aligned}
		\mathbb{H} =&i\hbar g(a^{\dagger 2}_A-a^2_A)/2 +i\hbar g(a^{\dagger 2}_B-a^2_B)/2\\&
	+\hbar \delta (a^{\dagger}_C a_C+a^{\dagger}_B a_B) + i\hbar\kappa_{APT}(a^{\dagger}_B a^{\dagger}_C-a_B a_C)\\&+ \hbar\kappa_{PT}(a^{\dagger}_A a_B+a^{\dagger}_B a_A),
	\end{aligned}
	\label{eq:hamiltonian}
\end{equation}
with $\dagger$ denoting Hermitian conjugate. In Eq.(\ref{eq:hamiltonian}), $\delta$ and $\kappa_{APT}$ represent the detuning and the nonlinear coupling coefficient, respectively, in the detuned non-degenerate parametric process (modes B and C). $g$ is the phase-sensitive amplification gain and  $\kappa_{PT}$ is the linear coupling coefficient. In principle, our theoretical model can be implemented on an integrated photonic platform. Specifically, two parallel periodically poled lithium niobate (PPLN) waveguides can be fabricated using existing nanofabrication technology. The poling periods of the two waveguides are independently engineered such that one waveguide supports degenerate parametric down-conversion for mode A, whereas the other employs a multi-period poling structure to simultaneously realize degenerate parametric down-conversion for mode B and detuned nondegenerate parametric down-conversion between modes B and C. The two nonlinear interactions are phase matched by distinct reciprocal lattice vectors provided by the multi-period poling structure, allowing them to coexist within the same waveguide\cite{trovatello2025quasi,zhang2020dual}. From the corresponding Heisenberg--Langevin equations ($\partial_z a_j=\frac{i}{\hbar}[\mathbb{H},a_j]$), the dynamical equations are
\begin{subequations}
	\begin{align}
		\frac{\partial a_A}{\partial z}&=g a^{\dagger}_A-i\kappa_{PT} a_B, \label{eq:aA_evolution} \\
		\frac{\partial a_B}{\partial z}&=g a^{\dagger}_B-i\kappa_{PT} a_A-i\delta a_B+\kappa_{APT} a^{\dagger}_C,\label{eq:aB_evolution}\\
		\frac{\partial a^{\dagger}_C}{\partial z}&=\kappa_{APT} a_B+i\delta a^{\dagger}_C.\label{eq:aC_evolution}
	\end{align}
\end{subequations}
Since the parametric amplification process is a sensitive amplification process, the commutation relation $\left[ a_j(z),a_j^{\dagger}(z)\right]=\left[ a_j(0),a_j^{\dagger}(0)\right]=1 $ can be maintained in the Heisenberg equations Eqs.(\ref{eq:aA_evolution})-(\ref{eq:aC_evolution}) without the need to include additional losses or Langevin noise. Eqs.(\ref{eq:aA_evolution})-(\ref{eq:aC_evolution}) can be transformed into the corresponding quadrature-operator evolutions using $q_j=(a^{\dagger}_j+a_j)/2$ and $p_j=i(a^{\dagger}_j-a_j)/2$ $(j=A,B,C)$, where $[q_j,p_j]=i/2$. That is,
\begin{equation}
		i\partial_{z}\hat{v}=H_{eff}\hat{v}
		, \label{eq:Evolution equation}
\end{equation}
with the column vector $\hat{v}=\left[q_A,q_B,q_C,p_A,p_B,p_C\right]^{T} $  and the non-Hermitian Hamiltonian $H_{eff}$, taking the form:
\begin{equation}
		H_{eff} = i\sigma_z\otimes\begin{bmatrix}
			g&0&0\\
			0&g&\kappa_{APT}\\
			0&\kappa_{APT}&0
		\end{bmatrix}-\sigma_y\otimes\begin{bmatrix}
		0&\kappa_{PT}&0\\
		\kappa_{PT}&\delta&0\\
		0&0&\delta
		\end{bmatrix}.\label{eq:H_qApBqC}
\end{equation}
Based on Eq.(\ref{eq:H_qApBqC}), $H_{eff}$ possesses a hidden non-Hermitian chiral symmetry \cite{kawabata2019symmetry} $UH_{eff}+H_{eff}U=0$ with $U=\sigma_{x}\otimes I_3$ and $I_3$ is a $3\times3$ identity matrix.
The eigenvalues of $H_{eff}$ are given by $\omega_{n,\pm}=\pm i\sqrt{\tilde{\omega}_{n,\pm}}$ and  $n\in\{1,2,3\}$, where $\tilde{\omega}_n$ can be obtained by solving the relevant characteristic polynomial equation $\tilde{\omega}_n^3+f_2\tilde{\omega}_n^2+f_1\tilde{\omega}_n+f_0=0$ with
\begin{subequations}
	\begin{align}
	&f_2=-2(g^2+\beta-\delta^2),\\
	&f_1=g^4+g^2(2\beta-3\delta^2)+(\delta^2-\beta)^2,\\
	&f_0=g^4\delta^2+\kappa_{PT}^4\delta^2-g^2(\kappa_{APT}^4-2\beta\delta^2+\delta^4).
    \end{align}
\end{subequations}
where $\beta=\kappa_{APT}^2-\kappa_{PT}^2$ quantifies the competition between the quadrature-PT and quadrature-APT. Figs~\ref{fig:scheme}(b) and (c) present the eigenspectra for the limiting cases of $\beta<0$ ($\{\kappa_{APT},\delta\}={0,0}$) and $\beta>0$ ($\{\kappa_{PT},g\}={0,0}$), respectively. These two limits recover the well-known eigenspectral features of quadrature-PT and quadrature-APT symmetries. A natural question then arises: how does the eigenspectrum evolve when the quadrature-PT and quadrature-APT interactions have equal strength, namely $\kappa_{PT}=\kappa_{APT}=\kappa$? More importantly, how do these two incompatible symmetries jointly shape the spectral properties of the system?

\begin{figure}[t]
	\centering\includegraphics[width=1\linewidth]{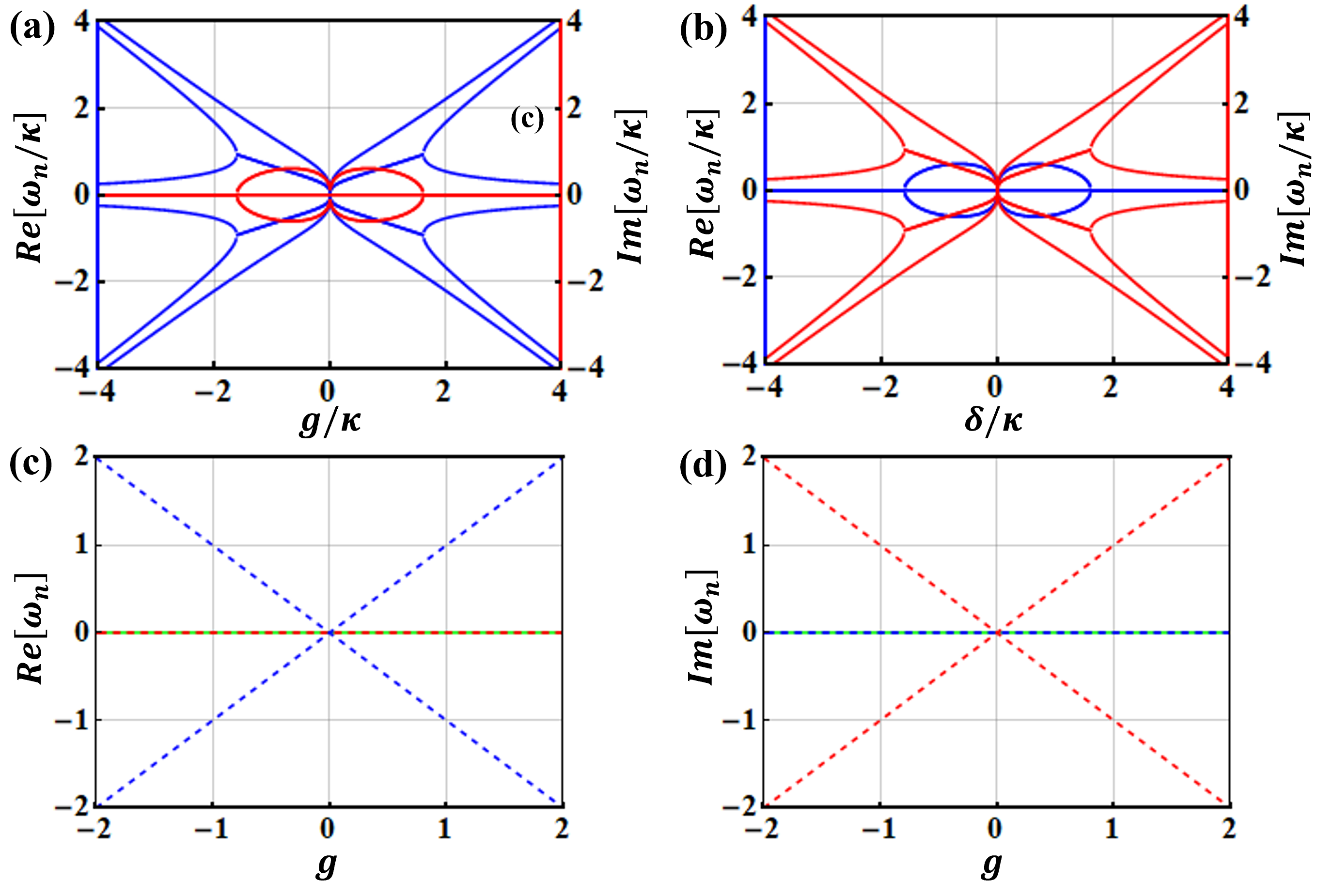}
	\caption{Evolution of the eigenspectrum under the combined action of quadrature-PT and quadrature-APT symmetries with $\kappa_{PT}=\kappa_{APT}=\kappa=1$. Real (blue solid) and imaginary (red solid) parts of the eigenvalues $\omega_n$ are shown for (a) $\delta=0$ and (b) $g=0$, illustrating the intermediate stages of spectral evolution. (c) and (d) show the real and imaginary parts of $\omega_n$, respectively, under the balanced condition $g=\delta$, where the Dirac eigenspectrum emerges. The dashed green, blue, and red lines represent the reference spectra $\omega_n=0$, $\omega_n=\pm g$, and $\omega_n=\pm ig$, respectively.}
	\label{fig:EP}
\end{figure} 

To address this question, Figs.~\ref{fig:EP}(a) and (b) show the eigenspectra for $\{\beta,\delta\}=\{0,0\}$ and $\{\beta,g\}=\{0,0\}$, respectively. When the PT and APT interactions are further balanced by setting $g=\delta$, the corresponding eigenspectra are shown in Figs.~\ref{fig:EP}(c) and (d). By comparing Figs.~\ref{fig:scheme}(b) and (c) with Fig.~\ref{fig:EP}, one can clearly trace the evolution of the eigenspectrum from two independent non-Hermitian spectral branches to a hybridized multi-branch structure, and finally to their complete coalescence. Correspondingly, the conventional non-Hermitian eigenvalue-splitting branches evolve into multiple intersecting and splitting branches before collapsing into a Dirac-point-like spectral structure.

Remarkably, this spectral degeneracy differs from that of both Hermitian Dirac points and conventional non-Hermitian exceptional points. At $g=\delta=0$, we can obtain,
\begin{equation}
\left\langle \psi_{\omega_n}^{L}\middle|\psi_{{\omega_n}}^{R}\right\rangle=0,\left\langle \psi_{\omega_{n,\pm}}^{R}\middle|\psi_{\omega_{n,\mp}}^{R}\right\rangle=1
\end{equation}
where $|\psi_n^{R}\rangle$ and $\langle\psi_n^{L}|$ are right and left eigenvectors of $H_{eff}$, satisfying
$H|\psi_n^{R}\rangle=\lambda_n|\psi_n^{R}\rangle$ and
\(\langle\psi_n^{L}|H=\lambda_n\langle\psi_n^{L}|\) (See Supplementary Materials for more details). We refer to this unique spectral structure as a \emph{non-Hermitian Dirac eigenspectrum}\cite{wu2025experimental}. Unlike conventional exceptional points, this characteristic spectrum is linear.

\section{Quantum Sensing and quantum Squeezing}
To uncover the quantum-fluctuation mechanism underlying the non-Hermitian Dirac eigenspectrum, we analyze the noise properties of collective three-mode quadratures measured via homodyne detection and compare them with the vacuum-noise limit. Specifically, we introduce the collective quadratures $Q_3=q_A-q_B-q_C$ and $P_3=p_A-p_B-p_C$ which satisfy the canonical commutation relation $\left[ Q_3,P_3\right] =i/2$. For concreteness, the three bosonic modes are assumed to be initially prepared in the coherent state $\left|\alpha_A,\alpha_B,\alpha_C\right\rangle$ with $\alpha_A=\alpha_B=\alpha_C=1$. Solving Eq.~(\ref{eq:Evolution equation}) under the balanced condition $g=\delta$ and $\kappa_{PT}=\kappa_{APT}=\kappa$, the variance of the collective quadrature is obtained as $\left\langle \Delta P_3^{2}\right\rangle=\frac{\mathcal{F}(g,\kappa,L)}{12g^{6}}$. The explicit expression of $\mathcal{F}$  and the results for $Q_3$ are given in the Supplemental Material.

\begin{figure}[t]
	\centering\includegraphics[width=1\linewidth]{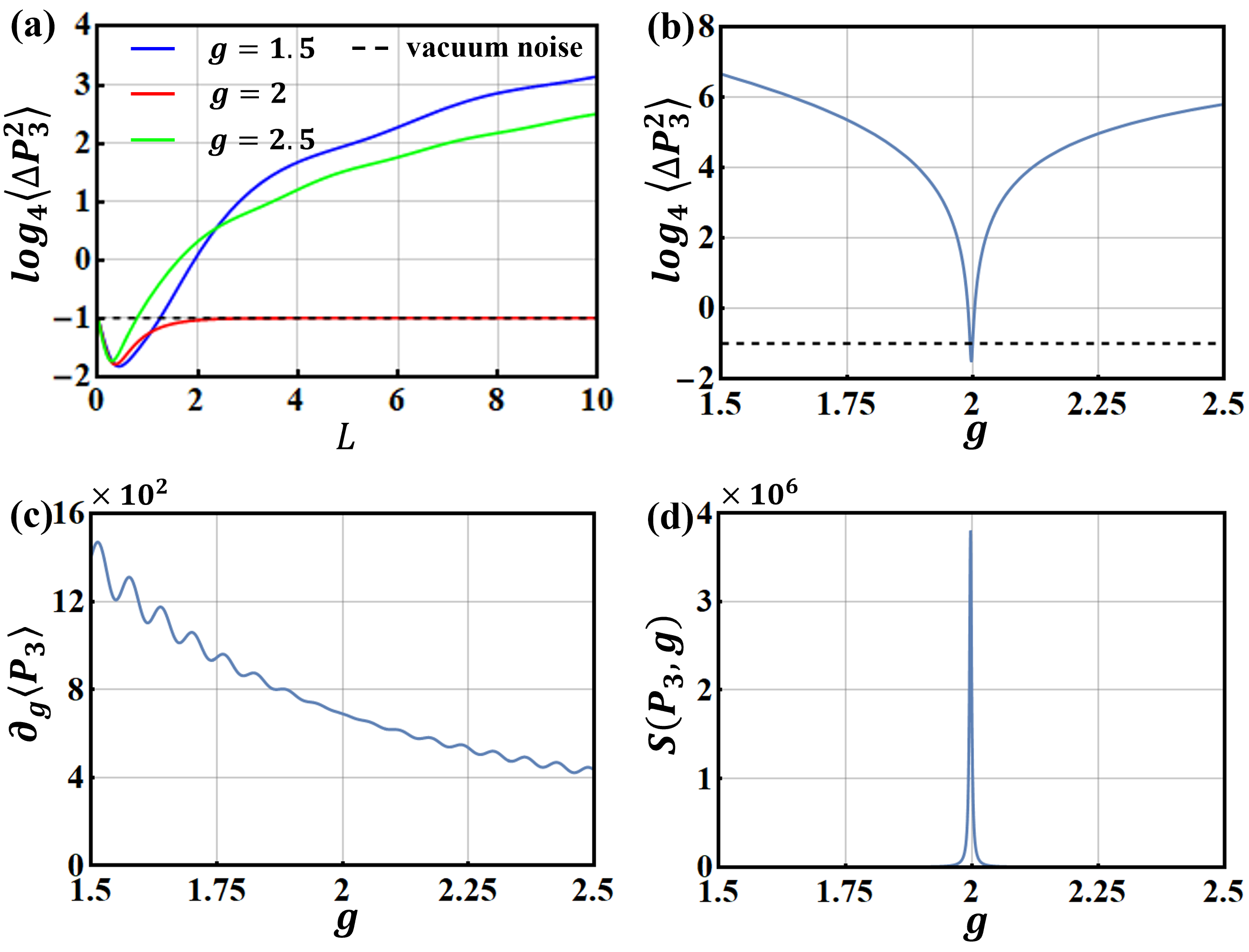}
	\caption{Chiral-symmetry-protected quantum fluctuations of the collective three-mode quadrature $P_3$. (a) Variance $\langle\Delta P_3^2\rangle$ as a function of the evolution length $L$. (b) Variance $\langle\Delta P_3^2\rangle$ as a function of the gain parameter $g$. The black dashed line ($\log_{4}\langle\Delta P_3^2\rangle=-1$) denotes the vacuum-noise level. A symmetry-protected vacuum-noise fixed point appears at $g=\kappa=2$, where the squeezing of $P_3$ becomes independent of the evolution length. (b)–(d) are evaluated with $L=100$ and $\kappa=2$. (c) Response sensitivity $\partial_g\langle P_3\rangle$. (d) Quantum sensing sensitivity $S(P_3,g)$. }
	\label{fig:sensing}
\end{figure}

The corresponding numerical results are presented in Figs.~\ref{fig:sensing}(a) and (b), where the variance is plotted as a function of $L$ and $g$, together with the vacuum-noise level (black dashed line). The results reveal a striking noise behavior originating from the interplay between parametric amplification and hidden chiral symmetry. Although the Dirac eigenspectrum exhibits no apparent spectral singularity [Figs.~\ref{fig:EP}(c) and (d)], the quantum fluctuations display a qualitatively distinct behavior at the hidden phase-transition point $g=\kappa$. In particular, the collective quadrature is always squeezed below the vacuum-noise limit at $g=\kappa$ and is exactly given by
\begin{equation}
	\begin{aligned}
		\left\langle \Delta P_3^{2}(g=\kappa)\right\rangle=
		\frac{1}{4}-\frac{2}{3}(e^{-\kappa L}-e^{-2\kappa L})\leqslant	\frac{1}{4}.
	\end{aligned}
\end{equation} 
Revisiting the evolution of the eigenspectrum presented in the previous section reveals that the transition of the Hamiltonian phase-transition point from $g(\delta)=\kappa_{PT}(\kappa_{APT})$ to $g=\kappa$ is not accidental, but instead originates from the protection of the hidden chiral symmetry.
Therefore, the hidden phase-transition point $g=\kappa$ is simultaneously a symmetry-protected vacuum-noise fixed (VF) point.

Since that symmetry-protected vacuum noise fixed point differs from the past situation where EPs amplify noise, this prompts us to explore its applications in quantum sensing. For $S(P_3,g)\equiv(\partial_{g}\langle P_3\rangle)^2/\left\langle \Delta P_3^{2}\right\rangle$, the response sensitivity $\partial_{g}\left\langle P_3 \right\rangle$ can be obtained,
\begin{align}
	\partial_{g}\langle P_3\rangle
	&=
	\frac{(g+\kappa)e^{-gL}\Bigl[6\kappa^{2}-g(g-2\kappa)(g+\kappa)L\Bigr]+\zeta}{\sqrt{3}\,g^{4}}.
	\label{eq:Rsensitivity}
\end{align}
with $\zeta=2\kappa\left(g^{2}+g\kappa-3\kappa^{2}\right)[\sin(gL)-\cos(gL)]+g(g-2\kappa)(g^2-\kappa^2)L[\sin(gL)+\cos(gL)]+2\kappa(g^{2}-2g\kappa-6\kappa^{2})+2g(g^{3}+g\kappa^{2}+4\kappa^{3})L$. The response sensitivity is plotted in Fig.~\ref{fig:sensing}(c) for $\{\kappa,L\}=\{2,100\}$. Unlike conventional EP-enhanced sensing, the symmetry-protected VFP does not exhibit any anomalous amplification of the response sensitivity. Nevertheless, as shown in Fig.~\ref{fig:sensing}(d), the sensing sensitivity reaches its maximum precisely at the symmetry-protected point.

To understand this behavior, we examine the asymptotic scaling near $g=\kappa$. In the limit $L\rightarrow\infty$, one finds $\lim_{\epsilon \to 0}[\partial_{g}\langle P_3\rangle]\sim O[1]$ whereas the excess quantum noise satisfies $\lim_{\epsilon \to 0}[\left\langle \Delta P_3^{2}\right\rangle-\frac{1}{4}]\sim C\epsilon^{2}+O[\epsilon^3]$
with $\epsilon=g-\kappa$. Therefore, the excess quantum fluctuations are quadratically suppressed in the vicinity of the symmetry-protected VFP, giving rise to the scaling law (The detailed derivation is provided in the Supplemental Material)
\begin{equation}
	\begin{aligned}
		S_{ex}(P_3,g)\equiv\frac{(\partial_{g}\langle P_3\rangle)^2}{\left\langle \Delta P_3^{2}\right\rangle-\frac{1}{4}}\sim \epsilon^{-2}.
	\end{aligned}
\end{equation}
These results establish a fundamentally different mechanism for quantum-enhanced sensing. In contrast to conventional EP-based approaches, where sensitivity enhancement relies on anomalous response amplification, the present scheme achieves sensing enhancement through symmetry-protected suppression of quantum fluctuations while maintaining a finite response sensitivity. This demonstrates that quantum sensing enhancement does not necessarily require anomalous response amplification. Instead, suppressing quantum fluctuations alone is sufficient to improve the achievable sensing performance. Our work therefore establishes noise suppression, rather than response amplification, as an alternative paradigm for non-Hermitian quantum sensing.

%Although the feasibility of quantum sensing based on non-Hermitian exceptional points remains under debate, there is a general consensus on why conventional non-Hermitian exceptional points fail to provide signal-to-noise ratio enhancement. In particular, the enhanced response sensitivity associated with exceptional points is inevitably accompanied by amplified noise fluctuations, which ultimately suppress the expected improvement in sensing performance. However, in conjunction with Eq. (\ref{eq:sensitivity}), this leads us to a different perspective. Instead of pursuing sensing enhancement through the amplification of response sensitivity, one may alternatively achieve quantum sensing enhancement by suppressing noise fluctuations. 
%In other words, reducing the noise floor can be just as effective as increasing the signal response for improving sensing performance.

%In the following, we verify this viewpoint using a three-mode quadrature homodyne detection scheme. Specifically, we define a set of three-mode quadratures, $q_3$ and $p_3$, satisfying $\left[ q_3,p_3\right] =i/2$.

\section{Conclusion}
In summary, we have proposed a fully quantum continuous-variable model formed by coupling a degenerate phase-sensitive amplifier and a nondegenerate four-wave-mixing process that unifies PT and APT symmetries within a single theoretical framework, which can be implemented in platforms such as integrated photonic systems. By exploiting the incompatibility between PT and APT symmetries, we uncover a non-Hermitian Dirac eigenspectrum and demonstrate that a symmetry-protected VFP can remain hidden within the Hamiltonian eigenspectrum. Through a systematic analysis of the quantum fluctuations, we identify a set of collective three-mode quadratures governed by this hidden fixed point.

Within the framework of quantum metrology, we show that these collective quadratures exhibit enhanced sensing performance at the symmetry-protected VFP. In sharp contrast to conventional EP-based sensing, where enhanced response sensitivity is inevitably accompanied by amplified quantum noise, the present mechanism achieves quantum sensing enhancement by suppressing quantum fluctuations while maintaining a finite response sensitivity. Consequently, the signal-to-noise ratio is improved without relying on anomalous response amplification.

Our results establish noise suppression as an alternative route toward non-Hermitian quantum-enhanced sensing and reveal a previously unexplored role of hidden symmetry-protected fixed points in quantum metrology. More broadly, this work enriches the interplay between symmetry and quantum fluctuations in non-Hermitian physics and opens new opportunities for symmetry-engineered quantum sensing.

\section{Acknowledgments}
%\emph{{Acknowledgments}.---}
This work is supported by the National Natural Science Foundation of China (62375089 and 62471188), the Innovation Program for Quantum Science and Technology (2021ZD0303401), the Quantum Science Strategic Project of Guangdong Province (GDZX2306004) and the Special Funds for the Cultivation of Guangdong College Students' Scientific and Technological Innovation ("Climbing Program" Special Funds) (Grant No.pdjh2025ak062).

\bibliography{bibliography}

\end{document}